\documentclass[twocolumn,superscriptaddress,aps,preprintnumbers,amsmath,amssymb,prl,nofootinbib]{revtex4-1}

\usepackage{graphicx}
\usepackage{epstopdf}
\usepackage{dcolumn}% Align table columns on decimal point
\usepackage{bm}% bold math
\usepackage{hyperref}
\usepackage{color}

\begin{document}
%%%%%%%%%%%%%%%%%%%%%%%%%%%%%%%%%%%%%%%%%%%

\def\a{\alpha}
\def\b{\beta}
\def\c{\varepsilon}
\def\d{\delta}
\def\e{\epsilon}
\def\f{\phi}
\def\g{\gamma}
\def\h{\theta}
\def\k{\kappa}
\def\l{\lambda}
\def\m{\mu}
\def\n{\nu}
\def\p{\psi}
\def\q{\partial}
\def\r{\rho}
\def\s{\sigma}
\def\t{\tau}
\def\u{\upsilon}
\def\v{\varphi}
\def\w{\omega}
\def\x{\xi}
\def\y{\eta}
\def\z{\zeta}
\def\D{\Delta}
\def\G{\Gamma}
\def\H{\Theta}
\def\L{\Lambda}
\def\F{\Phi}
\def\P{\Psi}
\def\S{\Sigma}

\def\o{\over}
\def\beq{\begin{eqnarray}}
\def\eeq{\end{eqnarray}}
\newcommand{\gsim}{ \mathop{}_{\textstyle \sim}^{\textstyle >} }
\newcommand{\lsim}{ \mathop{}_{\textstyle \sim}^{\textstyle <} }
\newcommand{\vev}[1]{ \left\langle {#1} \right\rangle }
\newcommand{\bra}[1]{ \langle {#1} | }
\newcommand{\ket}[1]{ | {#1} \rangle }
\newcommand{\EV}{ {\rm eV} }
\newcommand{\KEV}{ {\rm keV} }
\newcommand{\MEV}{ {\rm MeV} }
\newcommand{\GEV}{ {\rm GeV} }
\newcommand{\TEV}{ {\rm TeV} }
\def\diag{\mathop{\rm diag}\nolimits}
\def\Spin{\mathop{\rm Spin}}
\def\SO{\mathop{\rm SO}}
\def\O{\mathop{\rm O}}
\def\SU{\mathop{\rm SU}}
\def\U{\mathop{\rm U}}
\def\Sp{\mathop{\rm Sp}}
\def\SL{\mathop{\rm SL}}
\def\tr{\mathop{\rm tr}}

\def\IJMP{Int.~J.~Mod.~Phys. }
\def\MPL{Mod.~Phys.~Lett. }
\def\NP{Nucl.~Phys. }
\def\PL{Phys.~Lett. }
\def\PR{Phys.~Rev. }
\def\PRL{Phys.~Rev.~Lett. }
\def\PTP{Prog.~Theor.~Phys. }
\def\ZP{Z.~Phys. }

%%%%%%%%%%%%%%%%%%%%%%%%%%%%%%%%%%%%%%%%%%%%%%%%%%%%%%%%%%%%%%%

\title{
Non-Anomalous Discrete $R$-symmetry Decrees Three Generations
}

\author{Jason L.~Evans}
\affiliation{IPMU, University of Tokyo, Kashiwa, 277-8568, Japan}
\author{Masahiro Ibe}
\affiliation{IPMU, University of Tokyo, Kashiwa, 277-8568, Japan}
\affiliation{ICRR, University of Tokyo, Kashiwa, 277-8582, Japan}
\author{John Kehayias}
\affiliation{IPMU, University of Tokyo, Kashiwa, 277-8568, Japan}
\author{Tsutomu T.~Yanagida}
\affiliation{IPMU, University of Tokyo, Kashiwa, 277-8568, Japan}

\begin{abstract}
  We show that more than two generations of quarks and leptons are
  required to have an anomaly free discrete $R$-symmetry larger than
  $R$-parity, provided that the supersymmetric Standard Model can be
  minimally embedded into a grand unified theory.  This connects an
  explanation for the number of generations with seemingly unrelated
  problems like supersymmetry breaking, proton decay, the $\mu$
  problem, and the cosmological constant through a discrete
  $R$-symmetry.  We also show that three generations is uniquely
  required by a non-anomalous discrete $R$-symmetry in classes of
  grand unified theories such as the ones based on (semi-)simple gauge
  groups.
\end{abstract}

\date{\today}
\maketitle
\preprint{IPMU11-0186}
\preprint{ICRR-report-599-2011-16}

\section{Introduction}
Approximately 70 years ago I.~I.~Rabi famously quipped ``who ordered
that?''  in regards to the discovery of the second electron, i.e.~the
muon. Since that time the origin of multiple generations of quarks and
leptons has been a mystery. A partial answer to this question can be
found in the leptogenesis mechanism\,\cite{Fukugita:1986hr}.  In
leptogenesis, at least two generations of right-handed neutrinos are
required for $CP$-violation\,\cite{Frampton:2002qc}, an essential
ingredient in baryogenesis.  This solution, however, does not explain
the existence of the third generation\footnote{Explaining the origin
  of three generations has also been attempted in extra-dimensional
  models \cite{Witten:1982fp,Dobrescu:2001ae,Watari:2001qb}. The
  number of generations has also been related to discrete
  non-$R$-symmetries before \cite{Hinchliffe:1992ad,
  Mohapatra:2007vd}.}.

In this letter, we show that more than two generations of quarks and
leptons are necessary for an anomaly free discrete
$R$-symmetry\footnote{We assign $R$-charge $1$ to the Grassmann
  coordinate of superspace $\theta$, so that the superpotential has
  $R$-charge 2 modulo $N$.}, $\mathbb{Z}_{NR}$, of order $N>2$
\footnote{A $\mathbb{Z}_{2R}$ is not an $R$-symmetry; by including a
  Lorentz rotation, it is equivalent to (a non-$R$) $\mathbb{Z}_2$
  (see, for instance, \cite{dine_disr}).}. An $R$-symmetry is
important when considering model building and phenomenology with
supersymmetry: generic \cite{nelsonseiberg} and metastable
supersymmetry breaking \cite{iss}, proton decay \cite{protondecay},
and the $\mu$ problem (see, for instance, \cite{yanagidamu,
  dine_disr}) can all be solved by, or require, an
$R$-symmetry. Furthermore, without a discrete $R$-symmetry, a constant
term in the superpotential is allowed and expected to be of order the
Planck scale.  A large constant term in the superpotential
necessitates Planck scale supersymmetry breaking to cancel the large
cosmological constant. Therefore, low scale supersymmetric extensions
of the Standard Model have various difficulties which well motivate an
$R$-symmetry. Additionally, $R$-symmetries are motivated by
considering string theory, where they arise as ``leftover'' symmetries
from higher dimensional Lorentz groups. By considering a minimal
embedding into a grand unified theory (GUT), we show that this
discrete $R$-symmetry requires (at least) three generations to be
anomaly free.

Furthermore, if the GUT group is semi-simple, then considering
anomalies with $U(1)_Y$ shows that four and five generations are not
consistent with an anomaly-free discrete $R$-symmetry. More than five
generations will lead to a Landau pole in the theory; only three
generations is viable. Additionally, we will show that the discrete
$R$-symmetry forbids a $\mu$ term, successfully suppresses proton
decay, and is consistent with the see-saw mechanism for neutrino
masses.

\section{Anomaly Free Discrete $R$-symmetry}
Now, let us consider the anomaly free conditions of a discrete
$R$-symmetry.  (Notice that we are assuming that a discrete
$R$-symmetry stems from a gauged $R$-symmetry since no global
symmetries are expected in a quantum theory of
gravity\,\cite{Banks:2010zn}.)  In the following, we consider a class
of GUT models where each generation of the quark and lepton
supermultiplets are unified into a $\mathbf{10}$ and a $\mathbf{5}^*$
representation of $SU(5)_{\rm GUT}$.  Furthermore, we also assume that
there are no additional light degrees of freedom charged under the
supersymmetric Standard Model (SSM) gauge groups beyond the ones in
the SSM. In particular, we expect that the colored Higgs associated
with the SSM Higgs doublets have masses of order the GUT scale.

In this class of models, the anomaly free conditions from $SU(3)_c$
and $SU(2)_L$ gauge symmetries with the discrete $R$-symmetry give
\cite{Kurosawa:2001iq,Ibanez:1992ji},
\begin{align}
\label{eq:A3}
6 + n_g(3r_{10} &+ r_5-4)= 0\ ,\\
\label{eq:A2}
4+ n_g (3r_{10}+r_5-4) &+ (r_u+r_d-2) = 0 ,
\end{align}
respectively, where these equations are modulo $N$.  Here, $n_g$
denotes the number of generations, $r_{10},r_5,r_u,r_d$ are the
$R$-charges of the superfields ${\bf 10, 5^*}$, $H_u$, $H_d$,
respectively.  Notice that, in our discussion, the $R$-charges are
assumed to be generation independent.\footnote{ If the model has
  generation dependent $R$-charges, some of the quark/lepton mass
  matrix elements are overly suppressed due to the $R$-symmetry.} The
presence of Yukawa interactions constrains the $R$-charges
\begin{align}
\label{eq:Y1}
2r_{10} +r_u = 2\ , \quad
%\label{eq:Y2}
r_{10} + r_5+r_d =2\ ,
\end{align}
modulo $N$.\footnote{It should be noted that the presence of the
  Yukawa interactions of the GUT multiplets does not necessarily lead
  to the exact unifications of the Yukawa coupling constants at the
  GUT scale.  The GUT relations can be corrected by higher dimensional
  operators involving the fields of vanishing $R$-charges whose
  expectation values break the GUT symmetries.} By combining
Eqs.\,(\ref{eq:A3})--(\ref{eq:Y1}), the anomaly free conditions reduce
to
\begin{align}
\label{eq:R-Gen}
 6 - 4n_g &= 0~(\mathrm{mod}~N)\ ,\\
\label{eq:Higgs}
 r_u + r_d &=4~(\mathrm{mod}~N)\ .
\end{align}

The condition in Eq.\,(\ref{eq:R-Gen}) remarkably relates the number
of generations of quarks and leptons to the order of the discrete
$R$-symmetry.  Interestingly, this condition shows that no discrete
$R$-symmetry with $N>2$ is allowed for $n_g = 1,2$. In fact, three
generations or more are needed to allow $R$-symmetries with $N\ge 3$
(see also Table\,\ref{tab:R-Gen}).  Consequently, we find that anomaly
free discrete $R$-symmetries with $N>2$ require more than two
generations of quarks and leptons, provided that the SSM is minimally
embedded into GUT representations.

\begin{table}[t]
\begin{center}
\begin{tabular}{|c|c|}
\hline
$n_g$ & $\mathbb{Z}_{NR}$ \\
\hline
$1$ & $N=2,1$\\
\hline
$2$ & $N=2,1$ \\
\hline
$3$ & $N=6,3,2,1$\\
\hline
$4$ & $N=10,5,2,1$\\
\hline
$5$ & $N=14,7,2,1$\\
\hline
\end{tabular}
\end{center}
\caption{
The relation between the number of generations and the order of the discrete $R$-symmetry.
%We see that an $R$-symmetry with $N\ge 3$ is allowed only for $n_g\ge 3$.
}
\label{tab:R-Gen}
\end{table}%

Another interesting feature of the above conditions is that the
$R$-charges of $H_u$ and $H_d$ in Eq.\,(\ref{eq:Higgs}) forbid the
so-called $\mu$-term when $N\ge 3$.  Therefore, the $\mu$-term is
automatically small in this class of models.  The same arguments also
apply to the infamous dimension five proton decay operator, $\mathbf{
  10 \, 10 \, 10 \, 5^*}$\,\cite{Sakai:1981pk,Weinberg:1981wj}, since
the $R$-charge of this operator is
%\begin{eqnarray}
%\label{eq:proton}
$3r_{10} + r_{5} = 4 -(r_u+r_d) = 0 $
%\end{eqnarray}
modulo $N$.
%Here, we again used the conditions from the Yukawa interactions in Eq.\,(\ref{eq:Y1})
%and (\ref{eq:Y2}).
Thus, the dimension five proton decay operator is also automatically
suppressed by the discrete $R$-symmetry (see also \cite{protondecay,
  Lee:2011dya}).

One may wonder whether other anomaly free conditions such as
$\mathbb{Z}_{NR}U(1)_Y^2$, $\mathbb{Z}_{NR}^2U(1)_Y$,
$\mathbb{Z}_{NR}^3$, and $\mathbb{Z}_{NR}({\rm gravity})^2$, could
give further constraints. The anomalies that are not linear in
$\mathbb{Z}_{NR}$ ($\mathbb{Z}_{NR}^2U(1)_Y$, for instance) are
sensitive to questions about the UV structure of the theory and are
thus not useful here \cite{Banks:1991xj}.  The $\mathbb{Z}_{NR}({\rm
  gravity})^2$ anomaly condition is also not useful for constraining
the SSM $R$-symmetries since it depends on states not in the low
energy SSM spectrum (see, for instance, the discussion in
\cite{Lee:2011dya}).
The $\mathbb{Z}_{NR}U(1)_Y^2$ condition is also model dependent,
and hence, this anomaly is not useful 
without specifying the models.
(We will come back to this point later.)

\section{$R$-Invariant Grand Unified Theory}
So far, we have not discussed the mechanism which gives mass to the
colored Higgs, i.e.~the infamous doublet-triplet splitting problem.
It is, however, known to be difficult to realize doublet-triplet
splitting naturally in GUT models with a simple gauge group and only
the SSM matter content below the GUT scale.  In addition to this
naturalness problem, it was recently shown in
Ref.\,\cite{Fallbacher:2011xg} that a low scale SSM with discrete
$R$-symmetries having $N>2$ are not consistent with GUT models based
on a simple gauge group.

Here, we present an example of a GUT model where 
the doublet-triplet splitting can be naturally
realized\,\cite{Yanagida:1994vq,Hotta:1995cd,
  Izawa:1997he,Kitano:2001ie,Kitano:2005ez,Kitano:2006wm}
and is  based on a non-simple group.
   In fact, an
$R$-invariant GUT model of this class has been constructed in
Ref.\,\cite{Izawa:1997he} based on a $SU(5)_{\rm GUT}\times U(3)_{H}$
gauge symmetry (see also Ref.\,\cite{Kurosawa:2001iq}).  
In Table\,\ref{tab:GUT}, we show the $\mathbb{Z}_{6R}$-charge assignments,
which includes $R$-parity, of this GUT model which satisfies all the anomaly free conditions of
the previous section for $n_g = 3$.
We also find the model presented  in the table II with three generations
satisfy the anomaly free conditions,
$\mathbb{Z}_{6R}SU(5)_{\rm GUT}^2$ and $\mathbb{Z}_{6R}SU(3)_{\rm
  H}^2$.

Let us briefly review product group unification model in the Higgs
phase\footnote{In the Higgs phase, $U(1)_H$ is necessary since the GUT
  gauge group is broken by the vacuum expectation values of $Q$ and
  $\bar{Q}$.  The $U(1)_H$ gauge group in the present GUT model
  unfortunately destroys the automatic explanation of the charge
  quantization in the usual GUT model.  (Since the MSSM quarks/leptons
  are neutral under $U(1)_H$, the charge quantization of
  quarks/leptons still holds.)  }. In this model, no adjoint of
$SU(5)_{GUT}$ is required and the GUT gauge symmetry is broken by the
expectation values of the bi-fundamental fields $Q$ and $\bar{Q}$ in
Table \ref{tab:GUT} \cite{Yanagida:1994vq,Hotta:1995cd, Izawa:1997he},
\begin{eqnarray}
% \begin{array}{c}
\langle Q^\alpha_i \rangle = v \delta^\alpha_i 
%=
%\left(
%\begin{array}{c c c c c}
%v &0 &0 &0 &0 \\
%0 &v &0 &0 &0 \\
%0 &0 &v &0 &0 \\
%\end{array}
%\right)\ ,
%  \\
%  \\
\ , \quad
\langle \bar{Q}^i_\alpha \rangle= v \delta^i_\alpha 
%=
%\left(
%\begin{array}{c c c}
%v &0 &0 \\
%0 &v &0 \\
%0 &0 &v \\
%0 &0 &0 \\
%0 &0 &0 \\
%\end{array}
%\right) \ ,
% \end{array}
\ ,
\label{eq:pgutvev1}
\end{eqnarray}
where $v$ denotes a dimensionful parameter at the GUT scale
and the indices run $\alpha = 1-3$ and $i=1-5$. 
With the
above expectation values, the SM gauge groups are the
unbroken subgroups of $SU(5)_{\rm GUT} \times U(3)_H$.  Specifically,
$SU(3)_c$ and $U(1)_Y$ are the diagonal subgroups of
$SU(5)_{\rm GUT} \times U(3)_H$.  
The above expectation values are obtained as a
supersymmetric solution of the
superpotential\,\cite{Yanagida:1994vq,Hotta:1995cd, Izawa:1997he},
\begin{eqnarray}
% \begin{array}{r l}
   W &= &  \sqrt{2}\bar{Q}^i_\alpha \Phi^a(t^a)^\alpha_\beta Q^\beta_i
   +\sqrt{2}\bar{X}_\alpha \Phi^a(t^a)^\alpha_\beta X^\beta \\
&& +\sqrt{2}\bar{Q}^i_\alpha \Phi^0(t^0)^\alpha_\beta Q^\beta_i
  +\sqrt{2}\bar{X}_\alpha \Phi^0(t^0)^\alpha_\beta X^\beta
  -\sqrt{2}v^2\Phi^0\ , \nonumber
% \end{array}
 \label{eq:pgutpotential3}
 \end{eqnarray}
 where\footnote{ Here, we used the Gell-Mann matrix $t^a\,
   (a=1\dots8)$ with the normalizations, tr\,$[t^a t^b] =
   \delta^{ab}/2$ and $t^0 = {\bf 1}_{3\times 3} /\sqrt{6}$.} we have
 distinguished the octet and singlet $\Phi$ of $SU(3)_H$ by $\Phi^a$
 and $\Phi^0$, respectively.  We have omitted the coupling constants
 of each term.  $i$ denotes the $SU(5)_{\rm GUT}$ representations, and
 $\alpha, \beta$ the $SU(3)_H$ representations.

 With the above expectation value, the colored Higgs multiplets in $H$
 and $\bar{H}$ form Dirac mass terms with $\bar X$ and $X$
 via the superpotential
\begin{eqnarray}
 W =  \bar{H}_i Q^i_\alpha \bar{X}^\alpha +  H^i \bar{Q}^\alpha_i X_\alpha  \ .
\label{eq:pgutpotential2}
\end{eqnarray}
In this way, we can successfully realize doublet-triplet splitting while
forbidding the mass term of the Higgs doublets
(see Ref.\,\cite{Yanagida:1994vq,Hotta:1995cd, Izawa:1997he} for details).

\begin{table}[t]
\begin{center}
{\renewcommand\arraystretch{1.15}
\begin{tabular}{|c|c|c|c|c|c|c|c|c|c|}
\hline
& $\mathbf {10}$& $\mathbf {5^*}$& $H(\mathbf {5})$ & $\bar{H}(\mathbf { 5^*})$
&$Q(\mathbf {5})$ & $\bar{Q}(\mathbf {5^*})$&$X(\mathbf {1})$ & $\bar{X}(\mathbf { 1})$
& $\Phi({\mathbf 1})$\\
\hline
$U(3)_H$ & ${\mathbf 1}$ & ${\mathbf 1}$ & ${\mathbf 1}$ & ${\mathbf 1}$
& ${\mathbf  {3^*}}$ & ${\mathbf  {3}}$ & ${\mathbf  {3^*}}$& ${\mathbf  {3}}$ & ${\mathbf 8}+{\mathbf 1}$\\
\hline
$\mathbb{Z}_{6R}$  & $ -1$ &$3$ & $4$ & $0$ & $0$ & $0$ & $-2$ & $2$ &$2$\\
\hline
\end{tabular}
}
\end{center}
\caption{
The $R$-charge assignments of our model based on $SU(5)_{\rm GUT}\times U(3)_H$,
which is consistent with the see-saw mechanism (see Eq.\,(\ref{eq:Y3})).
The Higgs doublets are embedded into (anti-)fundamental representation
of $SU(5)_{\rm GUT}$ (i.e. $H_u \subset H({\mathbf { 5^*}})$
and $H_d \subset \bar{H}({\mathbf 5})$).
}
\label{tab:GUT}
\end{table}%

Finally, we comment on the gauge couplings. In the Higgs phase, the
low energy coupling $\alpha_{3_c}$ is given by $1/\alpha_{3_c} =
1/\alpha_5 + 1/\alpha_{3H}$ with a similar relationship between
hypercharge, $\alpha_{Y}$, and $\alpha_{1H}$ \cite{Izawa:1997he}. Here
$\alpha_5$, $\alpha_{3H}$, and $\alpha_{1H}$ are the gauge coupling
constants of $SU(5)_{GUT}$, $SU(3)_H$, and $U(1)_H$ respectively. 
Thus, the unification is not automatically realized in this model.
For
a strongly coupled $SU(3)_H$ and $U(1)_H$, however, 
we see that at the GUT scale the approximate GUT relation, 
i.e. $\alpha_{3_c}\simeq\alpha_2\simeq
\alpha_1$, still holds.

\section{Use of $U(1)_Y$ Anomalies}

As we have mentioned earlier, the anomaly free conditions involving
$\mathbb{Z}_{NR}$ and $U(1)_Y$ are model dependent and not as powerful
as the ones in Eqs.\,(\ref{eq:A3}) and \,(\ref{eq:A2}) on general
grounds.  With more specifications of the GUT models\footnote{The
  $\mathbb{Z}_{NR}U(1)_Y^2$ condition is only useful if $U(1)_Y$ is
  embedded in a non-abelian part of the GUT group. }, however, these anomaly conditions can
also play an important role.  Moreover, these conditions might be the
key to single out $n_g = 3$ out of $n_g > 2$.

For example, if we assume that the GUT group is semi-simple and the
normalization of the $U(1)_Y$ charges is as in the Standard Model
(SM), then the $\mathbb{Z}_{NR}U(1)_Y^2$ anomaly free condition can be
used. As a function of $n_g$ it is given by
\begin{equation}\label{eq:A1}
2\left(-10n_g+3\right) = 0~(\mathrm{mod}~N)\ .
\end{equation}
Substituting $n_g=3$ for each $N=6,3,2,1$, we find that the anomaly
free condition for $\mathbb{Z}_{NR}U(1)_Y^2$ is also satisfied. For
$n_g=4$, the allowed $R$ symmetries are $N=74, 37, 2$, which are not
consistent with the other anomaly conditions for $n_g=4$, with the
exception of the (non-$R$) $\mathbb{Z}_{2}$ (See Table I).  For
$n_g=5$, the allowed $R$ symmetries are $N=94,47, 2$ and again are not
consistent with the other anomaly conditions in Table I.  By
remembering that $n_g \ge 6$ leads to a Landau pole, we find that
three generations is uniquely required by a non-anomalous discrete
$R$-symmetry in Eqs\,(\ref{eq:A3}), (\ref{eq:A2}) and (\ref{eq:A1}).

It should be noted that the product group unification considered in
the previous section is not semi-simple, and hence, the uniqueness of
three generation does not hold and the model allows $n_g = 3,4,5$.
For the uniqueness of three generations, $U(3)_H$ is needed to be
successfully embedded into a larger semi-simple gauge group.

\section{Discussion}
Any discrete $R$-symmetry with $N>2$ should be spontaneously broken
down to $R$-parity at some scale much lower than the GUT or Planck
scale.  Such spontaneous breaking of exact discrete symmetries could
cause a cosmological domain wall problem\,\cite{Dine:2010eb}.  One
option for avoiding this domain wall problem is to assume that the
spontaneous breaking of the discrete $R$-symmetry occurs well before
inflation.  This leads to constraints on the Hubble constant during
inflation and the reheating temperature relative to the $R$-symmetry
breaking scale.  Another interesting possibility is a model where the
vacuum expectation value of the inflaton breaks the discrete $R$
symmetry\,\cite{Kumekawa:1994gx}.  In this class of models, the
flatness of the inflaton potential near the origin is naturally
explained by the $R$-symmetry\,\cite{Dine:2011ws}, while the domain wall
problem is avoided because the radius of the coherent domain of the
inflaton field is inflated to $e^{N_e}~(N_e \gtrsim 60)$ times larger
than the Hubble radius at the end of inflation.

In Ref.\,\cite{Lee:2011dya} (see also Ref.\,\cite{Babu:2002tx}),
anomaly cancellation of discrete $R$-symmetries was also studied, but
with the addition of the Green-Schwarz mechanism. Our approach is to
take the minimal setup and constraints, and as we stated above, we do
not include gravitational effects for the anomalies nor extra matter
content for cancellation. Including a Green-Schwarz mechanism
fundamentally changes the anomaly relations to be satisfied and the
resulting analysis. However, taking a minimal approach and requiring
the discrete $R$-symmetry to be anomaly free and unbroken (at the GUT
scale at least) alleviates proton decay problems without additional
assumptions or constraints on $R$-breaking when using mechanisms like
Green-Schwarz.

We also comment on the $R$-charge of the right-handed neutrinos, $r_1$,
which are essential for the see-saw mechanism\,\cite{seesaw}.
By including right-handed neutrinos,
we obtain two additional conditions on the $R$-charges;
\begin{align}
\label{eq:Y3}
 r_5 + r_1 + r_u = 2\ , \quad
 2 r_1= 2\ , ({\rm mod} N).
\end{align}
The first condition is due to the Yukawa interactions of
the right handed neutrinos, and the second condition is due to the
Majorana masses of the right-handed neutrinos.  By combining the
conditions in Eqs.\,(\ref{eq:A3})--(\ref{eq:Y1}) and (\ref{eq:Y3}), we find the $R$-charges in Table\,\ref{tab:GUT}.  Thus,
we find that the $\mathbb{Z}_{6R}$ symmetry is consistent with the
see-saw mechanism. However, we find that these charge assignments are not
consistent with $SO(10)$ unification since $r_{10}=r_{5}=r_{1}$ is not
satisfied, which was pointed out in Ref.\,\cite{Lee:2011dya}.

Lastly, we discuss some possible hints of this discrete $R$
symmetry. The discrete $R$ symmetry discussed above forbids a $\mu$
term. We can then add a singlet with a discrete $R$-charge of $4$ to
generate the $\mu$ term. Inclusion of this singlet leads to the NMSSM
like extension (but the model without a cubic term for the singlet in
the superpotential) discussed in \cite{Dine:2007xi}. Thus, a
relatively large Higgs boson mass and other NMSSM like singlet signals
would be quite suggestive of this framework.

In addition to the $\mu$ term naturally leading us to possible future
physical signals, we showed in the previous section that proton decay
is also addressed by this discrete $R$ symmetry; the phenomenological
aspects of a $\mathbb{Z}_{6R}$ symmetry are rich. 
To conclude, we have shown that a discrete $R$-symmetry (larger than
$\mathbb{Z}_2$) requires at least three generations of quarks and
leptons to be anomaly free, assuming a minimal embedding in a
GUT. This non-anomalous discrete $R$-symmetry is rich
phenomenologically, and a useful ingredient in model building.

\renewcommand{\bibsection} {

\end{document}
\section*{References}}
\begin{thebibliography}{99}

% \cite{Fukugita:1986hr}
\bibitem{Fukugita:1986hr}
    M.~Fukugita and T.~Yanagida,
    Phys.\ Lett.\ B {\bf 174}, 45 (1986).
    %%CITATION = PHLTA,B174,45;%%
%\cite{Frampton:2002qc}
\bibitem{Frampton:2002qc}
  P.~H.~Frampton, S.~L.~Glashow, T.~Yanagida,
  %``Cosmological sign of neutrino CP violation,''
  Phys.\ Lett.\  {\bf B548}, 119-121 (2002).
  [hep-ph/0208157];
%\cite{Endoh:2002wm}
%\bibitem{Endoh:2002wm}
  T.~Endoh, S.~Kaneko, S.~K.~Kang, T.~Morozumi, M.~Tanimoto,
  %``CP violation in neutrino oscillation and leptogenesis,''
  Phys.\ Rev.\ Lett.\  {\bf 89}, 231601 (2002).
  [hep-ph/0209020].

  %\cite{Witten:1982fp}
\bibitem{Witten:1982fp}
  E.~Witten,
  %``An SU(2) anomaly,''
  Phys.\ Lett.\  B {\bf 117}, 324 (1982).
  %%CITATION = PHLTA,B117,324;%%

%\cite{Dobrescu:2001ae}
\bibitem{Dobrescu:2001ae}
  B.~A.~Dobrescu and E.~Poppitz,
  %``Number of fermion generations derived from anomaly cancellation,''
  Phys.\ Rev.\ Lett.\  {\bf 87}, 031801 (2001)
  [arXiv:hep-ph/0102010].
  %%CITATION = PRLTA,87,031801;%%

%\cite{Watari:2001qb}
\bibitem{Watari:2001qb}
  T.~Watari and T.~Yanagida,
  %``Semi-simple unification on T(6)/Z(12) orientifold in the type IIB
  %supergravity,''
  Phys.\ Lett.\  B {\bf 520}, 322 (2001)
  [arXiv:hep-ph/0108057].
  %%CITATION = PHLTA,B520,322;%%

%\cite{Hinchliffe:1992ad}
\bibitem{Hinchliffe:1992ad}
  I.~Hinchliffe and T.~Kaeding,
  %``B+L violating couplings in the minimal supersymmetric Standard Model,''
  Phys.\ Rev.\ D {\bf 47}, 279 (1993).
  %%CITATION = PHRVA,D47,279;%%

%\cite{Mohapatra:2007vd}
\bibitem{Mohapatra:2007vd}
  R.~N.~Mohapatra and M.~Ratz,
  %``Gauged Discrete Symmetries and Proton Stability,''
  Phys.\ Rev.\ D {\bf 76}, 095003 (2007)
  [arXiv:0707.4070 [hep-ph]].
  %%CITATION = ARXIV:0707.4070;%%

%\cite{Dine:2009swa}
\bibitem{dine_disr}
  M.~Dine, J.~Kehayias,
  %``Discrete R Symmetries and Low Energy Supersymmetry,''
  Phys.\ Rev.\  {\bf D82}, 055014 (2010).
  [arXiv:0909.1615 [hep-ph]].

\bibitem{nelsonseiberg}
  % \cite{Nelson:1993nf}
  % \bibitem{Nelson:1993nf}
  A.~E.~Nelson and N.~Seiberg,
  % ``R symmetry breaking versus supersymmetry breaking,''
  Nucl.\ Phys.\ B {\bf 416}, 46 (1994) [arXiv:hep-ph/9309299].
  %% CITATION = NUPHA,B416,46;%%

\bibitem{iss}
  % \cite{Intriligator:2006dd}
  % \bibitem{Intriligator:2006dd}
  K.~A.~Intriligator, N.~Seiberg and D.~Shih,
  % ``Dynamical SUSY breaking in meta-stable vacua,''
  JHEP {\bf 0604}, 021 (2006) [arXiv:hep-th/0602239].
  %% CITATION = JHEPA,0604,021;%%
See also
%\cite{Dimopoulos:1997ww}
%\bibitem{Dimopoulos:1997ww} 
  S.~Dimopoulos, G.~R.~Dvali, R.~Rattazzi and G.~F.~Giudice,
  %``Dynamical soft terms with unbroken supersymmetry,''
  Nucl.\ Phys.\ B {\bf 510}, 12 (1998)
  [hep-ph/9705307].
  %%CITATION = HEP-PH/9705307;%%

\bibitem{protondecay}
  % \cite{Ibanez:1991wt}
  % \bibitem{Ibanez:1991wt}
  %L.~E.~Ibanez and G.~G.~Ross,
  % ``Should discrete symmetries be anomaly free?,''
  %% CITATION = CERN-TH-6000-91;%%
  % \cite{Ibanez:1991hv}
  % \bibitem{Ibanez:1991hv}
  L.~E.~Ibanez and G.~G.~Ross,
  % ``Discrete gauge symmetry anomalies,''
  Phys.\ Lett.\ B {\bf 260}, 291 (1991).
  %% CITATION = PHLTA,B260,291;%%
  % \cite{Ibanez:1991pr}
  % \bibitem{Ibanez:1991pr}
  L.~E.~Ibanez and G.~G.~Ross,
  % ``Discrete Gauge Symmetries And The Origin Of Baryon And Lepton
  % Number
  % Conservation In Supersymmetric Versions Of The Standard Model,''
  Nucl.\ Phys.\ B {\bf 368}, 3 (1992);
  %% CITATION = NUPHA,B368,3;%%
  % \cite{Kurosawa:2001iq}
  % \bibitem{Kurosawa:2001iq}
  K.~Kurosawa, N.~Maru and T.~Yanagida,
  % ``Nonanomalous R-symmetry in supersymmetric unified theories of
  % quarks and
  % leptons,''
  Phys.\ Lett.\ B {\bf 512}, 203 (2001) [arXiv:hep-ph/0105136];
  %% CITATION = PHLTA,B512,203;%%
  % \cite{Dreiner:2005rd}
  % \bibitem{Dreiner:2005rd}
  H.~K.~Dreiner, C.~Luhn and M.~Thormeier,
  % ``What is the discrete gauge symmetry of the MSSM?,''
  Phys.\ Rev.\ D {\bf 73}, 075007 (2006) [arXiv:hep-ph/0512163].
  %% CITATION = PHRVA,D73,075007;%%

\bibitem{yanagidamu}
  % \cite{Yanagida:1997yf}
  % \bibitem{Yanagida:1997yf}
  T.~Yanagida,
  % ``A solution to the mu problem in gauge-mediated
  % supersymmetry-breaking
  % models,''
  Phys.\ Lett.\ B {\bf 400}, 109 (1997) [arXiv:hep-ph/9701394].
  %% CITATION = PHLTA,B400,109;%%

%\cite{Banks:2010zn}
\bibitem{Banks:2010zn}
  T.~Banks and N.~Seiberg,
  %``Symmetries and Strings in Field Theory and Gravity,''
  Phys.\ Rev.\  D {\bf 83}, 084019 (2011)
  [arXiv:1011.5120 [hep-th]].
  %%CITATION = PHRVA,D83,084019;%%
  %\cite{Hellerman:2010fv}
%\bibitem{Hellerman:2010fv}
  S.~Hellerman and E.~Sharpe,
  %``Sums over topological sectors and quantization of Fayet-Iliopoulos parameters,''

  [arXiv:1012.5999 [hep-th]].

%\cite{Ibanez:1992ji}
\bibitem{Ibanez:1992ji}
  L.~E.~Ibanez,
  %``More about discrete gauge anomalies,''
  Nucl.\ Phys.\  {\bf B398}, 301-318 (1993).
  [hep-ph/9210211].
  \bibitem{Kurosawa:2001iq}
  K.~Kurosawa, N.~Maru and T.~Yanagida, in Ref. [\cite{protondecay}]

%\cite{Sakai:1981pk}
\bibitem{Sakai:1981pk}
  N.~Sakai and T.~Yanagida,
  %``Proton Decay In A Class Of Supersymmetric Grand Unified Models,''
  Nucl.\ Phys.\  B {\bf 197}, 533 (1982).
  %%CITATION = NUPHA,B197,533;%%

%\cite{Weinberg:1981wj}
\bibitem{Weinberg:1981wj}
  S.~Weinberg,
  %``Supersymmetry At Ordinary Energies. 1. Masses And Conservation Laws,''
  Phys.\ Rev.\  D {\bf 26}, 287 (1982).
  %%CITATION = PHRVA,D26,287;%%

%\cite{Banks:1991xj}
\bibitem{Banks:1991xj}
  T.~Banks and M.~Dine,
  %``Note on discrete gauge anomalies,''
  Phys.\ Rev.\ D {\bf 45}, 1424 (1992)
  [hep-th/9109045].
  %%CITATION = HEP-TH/9109045;%%

%\cite{Fallbacher:2011xg}
\bibitem{Fallbacher:2011xg}
  M.~Fallbacher, M.~Ratz and P.~K.~S.~Vaudrevange,
  %``No-go theorems for R symmetries in four-dimensional GUTs,''
  arXiv:1109.4797 [hep-ph].
  %%CITATION = ARXIV:1109.4797;%%
%\cite{Bachas:1995yt}
\bibitem{Bachas:1995yt}
  C.~Bachas, C.~Fabre and T.~Yanagida,
  %``Natural gauge-coupling unification at the string scale,''
  Phys.\ Lett.\  B {\bf 370}, 49 (1996)
  [arXiv:hep-th/9510094].
  %%CITATION = PHLTA,B370,49;%%

  %\cite{Yanagida:1994vq}
\bibitem{Yanagida:1994vq}
  T.~Yanagida,
  %``Naturally Light Higgs Doublets In The Supersymmetric Grand Unified Theories
  %With Dynamical Symmetry Breaking,''
  Phys.\ Lett.\  B {\bf 344}, 211 (1995)
  [arXiv:hep-ph/9409329];
  %%CITATION = PHLTA,B344,211;%%
  %\cite{Hotta:1995cd}
\bibitem{Hotta:1995cd}
  T.~Hotta, K.~I.~Izawa and T.~Yanagida,
  %``Dynamical Models for Light Higgs Doublets in Supersymmetric Grand Unified
  %Theories,''
  Phys.\ Rev.\  D {\bf 53}, 3913 (1996)
  [arXiv:hep-ph/9509201];
  %%CITATION = PHRVA,D53,3913;%%
  %\cite{Izawa:1997he}
\bibitem{Izawa:1997he}
  K.~I.~Izawa, T.~Yanagida,
  %``R invariant natural unification,''
  Prog.\ Theor.\ Phys.\  {\bf 97}, 913-920 (1997).
  [hep-ph/9703350].
  %\cite{Kitano:2001ie}
\bibitem{Kitano:2001ie}
  R.~Kitano and N.~Okada,
  %``Dynamical doublet-triplet Higgs mass splitting,''
  Phys.\ Rev.\  D {\bf 64}, 055010 (2001)
  [arXiv:hep-ph/0105220].
  %%CITATION = PHRVA,D64,055010;%%
  %\cite{Kitano:2005ez}
\bibitem{Kitano:2005ez}
  R.~Kitano and G.~D.~Kribs,
  %``Tripletless unification in the conformal window,''
  JHEP {\bf 0503}, 033 (2005)
  [arXiv:hep-ph/0501047].
  %%CITATION = JHEPA,0503,033;%%
  %\cite{Kitano:2006wm}
\bibitem{Kitano:2006wm}
  R.~Kitano,
  %``Dynamical GUT breaking and mu-term driven supersymmetry breaking,''
  Phys.\ Rev.\  D {\bf 74}, 115002 (2006)
  [arXiv:hep-ph/0606129].
  %%CITATION = PHRVA,D74,115002;%%
  %\cite{Dine:2010eb}
\bibitem{Dine:2010eb}
  M.~Dine, F.~Takahashi and T.~T.~Yanagida,
  %``Discrete R Symmetries and Domain Walls,''
  JHEP {\bf 1007}, 003 (2010)
  [arXiv:1005.3613 [hep-th]].
  %%CITATION = JHEPA,1007,003;%%


%\cite{Kumekawa:1994gx}
\bibitem{Kumekawa:1994gx}
  K.~Kumekawa, T.~Moroi and T.~Yanagida,
  %``Flat potential for inflaton with a discrete R invariance in supergravity,''
  Prog.\ Theor.\ Phys.\  {\bf 92}, 437 (1994)
  [arXiv:hep-ph/9405337].
  %%CITATION = PTPKA,92,437;%%

  %\cite{Dine:2011ws}
\bibitem{Dine:2011ws}
  M.~Dine and L.~Pack,
  %``Studies in Small Field Inflation,''
  arXiv:1109.2079 [hep-ph].
  %%CITATION = ARXIV:1109.2079;%%

\bibitem{seesaw}
 T. ~Yanagida, in Proceedings of the Workshop on Unified Theory and Baryon Number of the Universe, eds.  O. Sawada and A. Sugamoto (KEK, 1979) p.95;
 M. Gell- Mann, P. Ramond and R. Slansky, in Supergravity,
 eds. P. van Niewwenhuizen and D. Freedman (North Holland, Amsterdam,
 1979); 
 S.L. Glashow, in Quarks and Leptons, Carg\`{e}se 1979, eds. M. L\'{e}vy, et al., 
 (Plenum 1980 New York), p. 707;
 R.~N.~Mohapatra and G.~Senjanovic,
  %``Neutrino Mass and Spontaneous Parity Violation,''
  Phys.\ Rev.\ Lett.\  {\bf 44}, 912 (1980).
 See also %\cite{Minkowski:1977sc}
%\bibitem{Minkowski:1977sc}
  P.~Minkowski,
  %``mu --> e gamma at a Rate of One Out of 1-Billion Muon Decays?,''
  Phys.\ Lett.\  {\bf B67}, 421 (1977).
%\cite{Lee:2011dya}
\bibitem{Lee:2011dya}
  H.~M.~Lee, S.~Raby, M.~Ratz, G.~G.~Ross, R.~Schieren, K.~Schmidt-Hoberg and P.~K.~S.~Vaudrevange,
  %``Discrete R symmetries for the MSSM and its singlet extensions,''
  Nucl.\ Phys.\  B {\bf 850}, 1 (2011)
  [arXiv:1102.3595 [hep-ph]].
  %%CITATION = NUPHA,B850,1;%%

%\cite{Babu:2002tx}
\bibitem{Babu:2002tx}
  K.~S.~Babu, I.~Gogoladze, K.~Wang,
  %``Natural R parity, mu term, and fermion mass hierarchy from discrete gauge symmetries,''
  Nucl.\ Phys.\  {\bf B660}, 322-342 (2003).
  [hep-ph/0212245].
  
  
%\cite{Dine:2007xi}
\bibitem{Dine:2007xi}
  M.~Dine, N.~Seiberg and S.~Thomas,
  %``Higgs physics as a window beyond the MSSM (BMSSM),''
  Phys.\ Rev.\ D {\bf 76}, 095004 (2007)
  [arXiv:0707.0005 [hep-ph]].
  %%CITATION = ARXIV:0707.0005;%%

% \bibitem{OurWork}
% J.~L.~Evans, M.~Ibe, J.~Kehayias, and T.~T.~Yanagida in preparation.


\end{thebibliography}
